\documentclass[11pt,a4paper]{article}

\usepackage[english]{babel}
\usepackage[latin1]{inputenc}
\usepackage{enumerate} 
\usepackage{amsmath,amssymb,natbib,graphicx}

\newcommand{\genotypes}{\bc_1,\bc_2}
\newcommand{\bR}{\boldsymbol{R}}
\newcommand{\bc}{\boldsymbol{c}}

\newcommand{\cd}{\,|\,}
\newcommand{\den}{p}
\newcommand{\Pro}{p}

\DeclareMathOperator{\Exp}{\mathbb{E}}
\DeclareMathOperator{\Var}{\mathrm{Var}}

\title{Estimation of Parameters in DNA Mixture Analysis\footnote{This is an Author's Accepted Manuscript of an article published in \emph{Journal of Applied Statistics,}  July 2013 \copyright Taylor \& Francis, available online at: http://www.tandfonline.com/doi:10.1080/02664763.2013.817549.}
}

\author{Therese Graversen\\University of Oxford
 \and Steffen Lauritzen\\University of Oxford 
}

\begin{document}

\maketitle

\begin{abstract} In \cite{article:Gammamodel} a Bayesian network for
    analysis of mixed traces of DNA was presented using gamma
    distributions for modelling peak sizes in the electropherogram. It
    was demonstrated that the analysis was sensitive to the choice of
    a variance factor and hence this should be adapted to any new
    trace analysed. In the present paper we discuss how the variance
    parameter can be estimated by maximum likelihood to achieve this. The unknown
    proportions of DNA from each contributor can similarly be
    estimated by maximum likelihood jointly with the variance
    parameter. Furthermore we discuss how to incorporate prior
    knowledge about the parameters in a Bayesian analysis.  The
    proposed estimation methods are illustrated through a few examples
    of applications for calculating evidential value in casework and
    for mixture deconvolution.\\

 
\end{abstract}

\section{Introduction}

The \emph{DNA profile} of a person is the genetic information at a set
of chosen \emph{markers} across chromosomes. For each marker, a person
has two sequences of DNA called \emph{alleles}, and the pair of alleles
constitute the \emph{genotype} of that marker.  When a DNA trace is
analysed, it is first amplified by a \emph{polymerase chain reaction} (PCR), and then the allelic
composition of the trace is determined by electrophoresis.  For further details, see for example \cite{book:Butler}.

The size
of a peak on the corresponding electropherogram is roughly
proportional to the amount of the DNA in the trace of that particular allelic type. This
quantitative information about the allelic composition becomes
particularly important, when analysing mixed traces of DNA. 
 
We consider a model for analysing a mixed trace of DNA using
information about peak sizes for each present allele, as obtained from
the electropherogram for that trace. {There is now a substantial body
of literature on methods for exploiting this information in the
analysis and interpretation of DNA mixtures. Early attempts include for example
\cite{article:Evettdata,article:Perlindata,Clayton199855,Gill200690,Gill200891,wang2006least,article:PENDULUM}; none of these are fully model based but use various summaries of the peak size information to separate contributions into major and minor components.
In addition there are 
a number of articles using Bayesian networks or other variants of graphical models describing  the distribution of peak sizes, 
for example
\cite{CurranMCMC, rgc/sll/jm:fsi,article:Gammamodel,perlin:2011,cowell:etal:11,puch2012,cowell:etal:13}. The present
paper belongs to  the model based paradigm in the latter group of articles.}

An important parameter in the analysis of \cite{article:Gammamodel} was
a variance factor in the peak size distribution. There, a fixed value
was used for the variance factor across all markers and all cases,
although there were signs of sensitivity to the choice of this
value. It was therefore suggested that
this parameter should be adapted to each case.  In the present paper
we respond to the suggestion by developing methods for simultaneously
estimating the variance factor and the unknown mixture proportions for
a given trace.

\section{A Bayesian network for DNA mixture analysis}

Our  model is implemented as a Bayesian network along the lines
described in \cite{article:Gammamodel}. 
Below we summarize some of the
main features of the model and its use.

\subsection{The gamma model for peak sizes}

For each allele present in the mixture the size of the corresponding
peak is observed; the size is represented by the peak area or peak height and 
possibly corrected for preferential amplification. A key assumption is
that the peak size is roughly proportional to the pre-amplification
amount of the corresponding allele \citep{Clayton199855}.

We are adopting the gamma model described in \cite{article:Gammamodel}
and partly justified in \cite{Cowell2009193}. The model assumes a
known number of contributors, and for technical simplicity we consider
here only cases with two contributors and do not allow for artefacts
such as stutter and drop-out.  We also assume that the
pre-amplification proportions of DNA from the two contributors is
constant across markers.  We represent the proportion of DNA
originating from one of the contributors by $\theta$; $\theta$ is then
a number between 0 and 1.

In \cite{article:Gammamodel} it is assumed that, for fixed genotypes
of the contributors and a fixed mixture proportion, the peak size
$W_a$ of allele $a$ at a given marker is independent of peak sizes of
other alleles and gamma distributed as
\begin{equation}\label{eq:gamma}
  W_a  \sim \Gamma(\beta\mu_a, \eta),
\end{equation}
where $\eta$ is a scale parameter, 
  $\mu_a  = \{\theta n^1_{a} + (1-\theta)n^2_{a}\}/2$,
and $n^1_{a}$ and $n^2_{a}$ denote the number of alleles of type $a$
at a given marker in the genotype of each contributor. Thus, for
example, if the first contributor has genotype $(13,15)$ and
contributed 40\% of the DNA, and the second contributor has genotype
$(15,15)$, then $n^1_{13}=n^1_{15}=1$, $n^2_{15}=2$, and all other
$n_a^i$-s are equal to zero. Hence, in this case
\begin{equation}\label{eq:mixex}\mu_{13}=\theta/2=0.20,\quad\mu_{15}=\{\theta+2(1-\theta)\}/2=1-\theta/2=0.80.
\end{equation}

At each marker the peak
sizes $(W_1,\ldots,W_A)$ are scaled by their sum such that
the resulting \emph{relative peak sizes} $(R_1,\ldots,
R_{A})$ add up to 1.
We let $\bR$ denote the total set of observed relative peak
sizes for all markers. Then $\bR$ follows a Dirichlet distribution.

The relative peak sizes are independent between markers and each 
$R_a$ follows a beta distribution with mean and variance given as
$$  \Exp R_a = \mu_a,\quad
\Var R_a = \sigma^2\mu_a(1 - \mu_a).$$ where we have let $\sigma =
1/\sqrt{\beta + 1}$.  Hence $\mu_a$ is the mean (relative) peak size
for allele $a$ so, for example, in the mixture (\ref{eq:mixex}) above
we would expect the peak at allele $15$ to be about four times as
large as that at $13$.  Also, $\sigma$ is a measure of the
\emph{generic peak imbalance}: For a single heterozygous contributor
with allele $a$ we have $\mu_a=1/2$ and therefore expect two peaks of
same size; the coefficient of variation for one such peak being
$$
  \frac{\sqrt{\Var R_a}}{\Exp R_a} = \frac{\sqrt{\sigma^2
      \tfrac{1}{2}\left(1-\tfrac{1}{2}\right)}}{\tfrac{1}{2}} =
  \sigma,
$$
i.e.\ if $\sigma=0.07$, say, the standard deviation of such a relative peak area is 7\%.
The parameter $\beta$ is related to the heterozygote balance ($Hb$) as
described in \cite{article:PENDULUM}, i.e.\ the ratio between the peak
sizes for the two alleles. The gamma model implies that $Hb$ is
$F(\beta, \beta)$-distributed. For a case where $\sigma=0.07$ we get
$\beta = 203.08$ and a 95\% prediction interval for $Hb$ would be
$0.759\le Hb \le 1.318$ which conforms well with previous findings
\citep{article:PENDULUM,Gill200690,Gill200891}.

\subsection{DNA mixture analysis}\label{sec:peakinfo}
Based on the relative peak areas and the Bayesian network, two key
questions can be addressed: a \emph{mixture deconvolution} which
attempts to determine the DNA profiles of the unknown contributors to
the mixture, and the calculation of an \emph{evidential value} for the
comparison of specific hypotheses concerning the composition of the
observed mixture.
\subsubsection{Mixture deconvolution}
The DNA profiles of the contributors to a mixture can be predicted by
a ranked list of most probable profile pairs $(\genotypes)$ based on the information in
the peak sizes, i.e.\ ranking these  according
to their probabilities $\Pro(\genotypes\cd \bR,\theta,\sigma)$. Note
that both $\theta$ and $\sigma$ are unknown and therefore need to be
estimated.

\subsubsection{Evidential value}
Suppose we have a reference profile from an individual which we shall
term the \emph{suspect} and wish to compare two specific hypotheses
$H_p$ and $H_d$, entertained by the prosecution and defence, for
example
  \begin{enumerate}
  \item[]  $H_p$: ``The suspect and one unknown individual has contributed to the trace''
    \item[]
    $H_d$: ``Two unknown individuals have contributed to the trace''.
  \end{enumerate}
  We consider contributors to be unrelated and the unknown individuals
  drawn at random from a specific population.  To assess the strength
  of the evidence we wish to calculate the likelihood ratio $LR$ of
  $H_p$ against $H_d$:
  \begin{equation}
    \label{eq:LR}
    LR  = \frac{\den(\bR\cd H_p,\theta,\sigma)}{\den(\bR\cd H_d,\theta,\sigma)},
  \end{equation}
where we again note the dependency of this ratio on the unknown parameters $\theta$ and $\sigma$.

\subsection{Data and software}

We illustrate the methods using relative peak sizes from two
mixtures with partial or complete knowledge of the contributors also used in
\cite{article:Gammamodel},
denoted the \emph{Evett} \citep{article:Evettdata} and \emph{Perlin}
\citep{article:Perlindata} data respectively. The peak sizes are adjusted for
preferential amplification by scaling the areas by the repeat number
for the corresponding allele. The Evett data
(Table~\ref{tab:Evettdata}) consists of the relative peak sizes from
a mixture in 10:1 ratio with a known profile for the main contributor.
The Perlin data (Table~\ref{tab:Perlindata}) are from a 7:3 ratio
mixture with two known contributors.

\begin{table}
  \centering
  \begin{minipage}[t]{0.45\linewidth}
      \caption{Evett data. The person with DNA profile $\bc_1$ is the major
        contributor. The profile for the minor contributor is unknown. \label{tab:Evettdata}} 
        \vspace{\baselineskip}  
        \begin{center}
        {\footnotesize
      \begin{tabular}{lcrc}
        Marker & Allele & $R$ & $\bc_1$\\ 
        \hline
        D8 & 10 & 0.4347 & 10 \\ 
        & 11 & 0.0285 &  \\ 
        & 14 & 0.5368 & 14 \\ 
        \hline
        D18 & 13 & 0.8871 & 13 \\ 
        & 16 & 0.0536 &  \\ 
        & 17 & 0.0592 &  \\ 
        \hline
        D21 & 59 & 0.0525 &  \\ 
        & 65 & 0.0676 &  \\ 
        & 67 & 0.4284 & 67 \\ 
        & 70 & 0.4515 & 70 \\ 
        \hline
        FGA & 21 & 0.5699 & 21 \\ 
        & 22 & 0.3908 & 22 \\ 
        & 23 & 0.0393 &  \\ 
        \hline
        TH01 & 8 & 0.4015 & 8 \\ 
        & 9.3 & 0.5985 & 9.3 \\ 
        \hline
        VWA & 16 & 0.4170 & 16 \\ 
        & 17 & 0.0884 &  \\ 
        & 18 & 0.4747 & 18 \\ 
        & 19 & 0.0199 &  \\ 
        \hline
      \end{tabular}
    }
    \end{center}
  \end{minipage}
  \hfill
  \begin{minipage}[t]{0.45\textwidth}
    {\footnotesize
      \caption{Perlin data. The person with DNA profile $\bc_1$ is the major contributor.}
      \label{tab:Perlindata}
      \vspace{\baselineskip}  
      \begin{center}
      \begin{tabular}{lcrcc}
        Marker & Allele & $R$ & $\bc_1$ & $\bc_2$\\ 
        \hline
        D2 & 16 & 0.1339 &  & 16 \\ 
        & 18 & 0.2992 & 18 &  \\ 
        & 20 & 0.1947 &  & 20 \\ 
        & 21 & 0.3722 & 21 &  \\ 
        \hline
        D3 & 14 & 0.5010 & 14 & 14 \\ 
        & 15 & 0.4990 & 15 & 15 \\ 
        \hline
        D8 & 9 & 0.2832 & 9 &  \\ 
        & 12 & 0.1426 &  & 12 \\ 
        & 13 & 0.3829 & 13 &  \\ 
        & 14 & 0.1913 &  & 14 \\ 
        \hline
        D16 & 11 & 0.6801 & 11 &  \\ 
        & 13 & 0.1607 &  & 13 \\ 
        & 14 & 0.1593 &  & 14 \\ 
        \hline
        D18 & 12 & 0.1504 &  & 12 \\ 
        & 13 & 0.3290 & 13 &  \\ 
        & 14 & 0.3443 & 14 &  \\ 
        & 17 & 0.1764 &  & 17 \\ 
        \hline
        D19 & 12.2 & 0.3109 & 12.2 &  \\ 
        & 14 & 0.3092 &  & 14 \\ 
        & 15 & 0.3799 & 15 &  \\ 
        \hline
        D21 & 27 & 0.1289 &  & 27 \\ 
        & 29 & 0.3913 & 29 &  \\ 
        & 30 & 0.4798 & 30 & 30 \\ 
        \hline
        FGA & 19 & 0.4621 & 19 & 19 \\ 
        & 24 & 0.1561 &  & 24 \\ 
        & 25.2 & 0.3817 & 25.2 &  \\ 
        \hline
        TH01 & 6 & 0.1268 &  & 6 \\ 
        & 7 & 0.4691 & 7 & 7 \\ 
        & 9 & 0.4041 & 9 &  \\ 
        \hline
        VWA & 17 & 0.7265 & 17 &  \\ 
        & 18 & 0.2735 &  & 18 \\ 
        \hline
      \end{tabular}
    \end{center} 
    }
  \end{minipage}
\end{table}

We follow \cite{article:Gammamodel} and use allele frequencies for the
US Caucasian population as given in \cite{butler:etal:03}. One of the
observed alleles, allele 25.2 at marker FGA, found in the Perlin
dataset was not present in the database, so the two known profiles
under study were added to the database and allele frequencies updated
accordingly.

We have used the software \texttt{R} \citep{manual:R} and
\texttt{HUGIN} \citep{manual:HUGINapi} for calculations in the
examples.  Through the \texttt{R}-package \texttt{RHugin}
\citep{manual:RHugin} it has been possible to perform all computations
from within \texttt{R} and hence take direct advantage of the
statistical tools available in \texttt{R} as well as those provided by
\texttt{HUGIN} for efficient computation in Bayesian networks.

\section{Methods for parameter estimation}

We now turn to the problem of estimating the unknown quantities $\sigma$
and $\theta$. We discuss three methods for doing so. 
\begin{enumerate}[(i)]
\item In the first method we proceed as in \cite{article:Gammamodel} and include $\theta$ in discretised form directly as a
  node in the Bayesian network with a
  uniform distribution. Instead of fixing a value $\sigma$ in advance we 
  estimate $\sigma$ by the method of maximum likelihood based on the case data at hand;
\item The second method treats also $\theta$ as a fixed and unknown
  parameter  and
  then estimates both $\sigma$ and $\theta$ by maximum likelihood;\item
  A third approach exploits prior information on both $\sigma$ and
  $\theta$ to perform a Bayesian analysis using Markov chain
  Monte Carlo methods \citep{gilks:richardson:spiegelhalter:96a}.
\end{enumerate}

\subsection{Maximum likelihood estimation of $\sigma$}\label{sec:sigmaestimation}
The likelihood function for $\sigma$  is  obtained by averaging out over all possible compositions $({\bc}_1, \bc_2, \theta)$ of the mixture: 
\begin{alignat*}{3}
  L(\sigma)= \den(\bR\cd H,\sigma) =  \sum_{\genotypes,\theta} \left\{\prod_m
      \den(\bR^m\cd \bc_1^m,\bc^m_2, \theta,\sigma)\right\}\den(\genotypes,\theta\cd H)
\end{alignat*}
where $\bR^m$, $\bc_1^m$ and $\bc_2^m$ are relative peak sizes and genotypes for each marker $m$ and $H$ denotes a specific hypothesis under consideration.
Direct computation of the likelihood function using this expression is not feasible as
the number of possible mixture compositions $(\genotypes,\theta)$ typically is overwhelming. However, the exact value of
$L(\sigma)$ can be obtained  as the normalising
constant from propagation of likelihood evidence in the Bayesian network 
which is what we have used here. We omit the technical details. 

\begin{figure}
  \centering
 \includegraphics{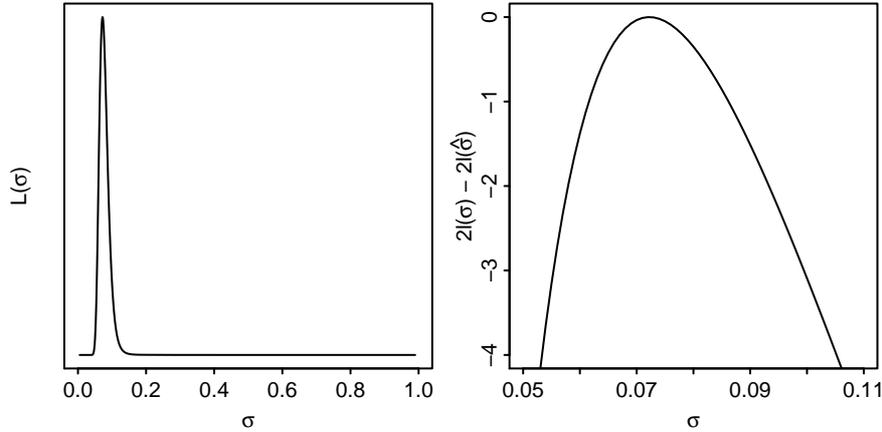}
  \caption{The likelihood function $L(\sigma)$ and its logarithm
    $\ell(\sigma) = \log L(\sigma)$ for the Perlin data and
    a scenario of two unknown contributors.}
  \label{fig:PerlinBetaLoglikelihood}
\end{figure}

Figure~\ref{fig:PerlinBetaLoglikelihood} shows the likelihood function
and its logarithm for the Perlin data when considering both
contributors unknown. The likelihood function can for example be
maximised using a general numeric algorithm for maximising a real
function. The likelihood function for the Perlin data has a maximum at
$\hat \sigma = 0.0722$, indicating a peak imbalance about 7\%.
The shape of $\ell(\sigma)$ around its maximum indicates that the
uncertainty of the MLE can reasonably be based on asymptotic normality
using the second derivative of the log-likelihood function as
$$\Var(\hat\sigma)\approx - 1/
\ell''(\hat\sigma).$$ This quantity can again be found by numerical
derivation; combined with using the normalising constant from 
propagation in the Bayesian network for exact computation of $\ell$, this is an extremely fast method.
Using this method for the Perlin data we obtain a 
99\% confidence interval for $\sigma$ of $(0.0441, 0.1003)$. In
comparison, \cite{article:Gammamodel} used a value of $\sigma^2 =
0.01$ corresponding to $\sigma = 0.1$, which is just inside the
confidence interval calculated.

\subsection{Maximum likelihood estimation of $\sigma$ and $\theta$}
\label{sec:sigmathetaestimation}
In contrast to the previous section we now also consider $\theta$ as a parameter and thus estimate both $\theta$ and $\sigma$ by maximising  the likelihood function
\begin{alignat*}{3}
  L(\theta,\sigma)= \den(\bR\cd H,\theta,\sigma) &=  \sum_{\genotypes} \left\{\prod_m
      \den(\bR^m\cd \bc_1^m,\bc^m_2, \theta,\sigma)\right\}p(\genotypes\cd H)\\
      &=\prod_m\left\{\sum_{\bc_1^m,\bc_2^m} 
      \den(\bR^m\cd \bc_1^m,\bc^m_2, \theta,\sigma)p(\bc_1^m,\bc_2^m\cd H)\right\}.
\end{alignat*}

To obtain the last equality we have used that when both of $\theta$
and $\sigma$ are fixed, the genotypes and peak sizes are all
independent between markers. The internal sums in the last expression
can be calculated as they stand, as each only involves genotypes at a
single marker.  Alternatively, $L(\theta,\sigma)$ can also here be
found from the normalising constant from propagation of the likelihood
evidence.
  
The asymptotic covariance matrix for the estimates is obtained from
the second derivatives of the log-likelihood function as before. Again we have maximised the likelihood function and found its derivatives by numerical methods.

\begin{figure}
  \centering
  \includegraphics{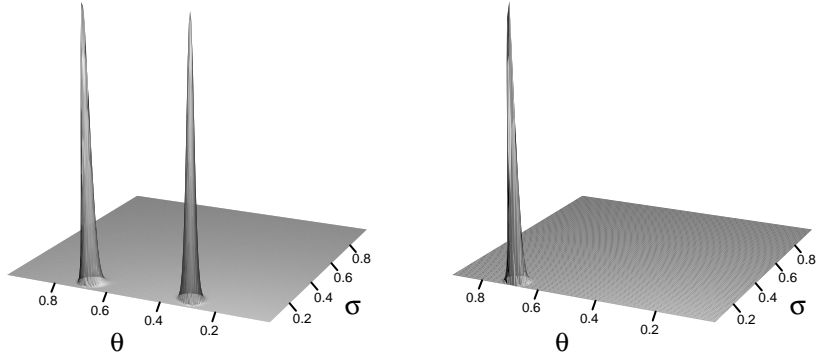}
  \caption{The likelihood function
    $L(\sigma, \theta) = \den(\bR;\sigma, \theta)$ for the Perlin data
    with two unknown contributors (left). To the right the likelihood
    function after specifying the DNA profiles for two
    contributors.}
  \label{fig:PerlinTwoParL}
\end{figure}

In the left-hand panel of Figure~\ref{fig:PerlinTwoParL} we see the
likelihood function for the Perlin data obtained in the case with two
unknown contributors. Unsurprisingly, the likelihood is symmetrical
around $\theta = 0.5$, because the labelling of contributors is arbitrary. 
The right-hand panel of Figure~\ref{fig:PerlinTwoParL} shows the likelihood function when the DNA profiles of both contributors are specified; the
likelihood function picks up which of the two contributors is the
major contributor and again correctly estimates the proportion of DNA
from this contributor to be around 0.7. 

The maximum likelihood estimates for $\sigma$ and $\theta$ are displayed in Table~\ref{tab:mleresults}. The estimates $\hat\sigma$ and
 $\hat\theta$ are close to being independent with asymptotic
correlations in the three situations for the Perlin data being -0.195,
-0.042, and -0.042. For the Evett data it is -0.160 in both
situations. For both data sets the estimated mixture proportions
$\theta$ are remarkably close to the proportions used for constructing
the DNA mixture. In contrast to the model using a uniformly
distributed $\theta$, the Perlin data does not quite support the use of
$\sigma = 0.1$ although it is not far off.

For the Perlin data, if we include genotypes of the minor contributor
as a potential contributor we get better estimates of the parameters
which is reflected in the narrower confidence intervals.  When further
including the DNA profiles of both contributors as known, the
estimates do not change at all.  For the Evett dataset, specifying
genetic information on a potential contributor barely changes the
estimates. 

For the Perlin data --- where $\sigma \approx 0.07$ --- a 95\% prediction
interval for the heterozygote balance $Hb$ is $0.759\le Hb \le
1.318$. For the Evett case the generic peak imbalance is a bit
higher, resulting in a slightly wider range of expected heterozygote
balance, $0.687 \le Hb \le 1.456$.  Note that for both the Perlin and
the Evett data the model leads to heterozygote balances that comply
with the recommendation in \cite{article:PENDULUM}.

\begin{table}

  \caption{\label{tab:mleresults} Joint maximum likelihood estimates of the mixture ratio and peak imbalance. The estimates of $\theta$ reflect the ratios used 
    for constructing the data; a 7:3 ratio for the Perlin data, and a 10:1 ratio for the Evett data.}
 
 \vspace{\baselineskip}
  \centering
  
{\footnotesize
  \begin{tabular}{lcccc}
\multicolumn{5}{c}{Perlin data}\\
   Genotype information & $\hat\sigma$ & 99\% CI & $\hat\theta$ & 99\% CI \\ 
    \hline
    Both contributors unknown & 0.070 & (0.040, 0.100) & 0.692 & (0.658, 0.727) \\ 
    One known potential contributor & 0.067 & (0.041, 0.094) & 0.696 & (0.666, 0.725) \\ 
    Both contributors known & 0.067 & (0.041, 0.094) & 0.696 & (0.667, 0.725) \\ 
    \hline
    &&&&\\
    \multicolumn{5}{c}{Evett data}\\
 Genotype information & $\hat\sigma$ & 99\% CI & $\hat\theta$ & 99\% CI \\ 
    \hline
    Both contributors unknown & 0.096 & (0.044, 0.147) & 0.895 & (0.858, 0.932) \\ 
    One known potential contributor & 0.096 & (0.044, 0.147) & 0.895 & (0.858, 0.932) \\ 
    \hline
  \end{tabular}
}  
\end{table}

\subsection{Including prior information about $\sigma$ and $\theta$}\label{sec:Bayesian}

In Section~\ref{sec:sigmaestimation} it was seen that the DNA mixture
can be modelled conditionally on the observed relative peak sizes
for a fixed $\sigma$ and a uniform distribution for $\theta$. We now
explain how to combine this model with prior information about the
variability on $\sigma$ to perform Bayesian inference in the model.

It is possible to simulate from $\Pro(\genotypes,\sigma, \theta \cd H,\bR)$, for example
by using a  Gibbs sampler which alternates between
\begin{enumerate}
\item sampling a pair $(\genotypes,\theta)$ of complete configurations
  of genotypes and mixture proportion given the current value of
  $\sigma$ and observed relative peak sizes;
\item sampling  $\sigma$ given the pair of DNA profiles sampled in the above step,
  the sampled mixture proportion, and the observed relative peak sizes.
\end{enumerate}

The first step is performed by sampling from the Bayesian network
model of the DNA mixture after including likelihood evidence using
$\sigma$ and $\bR$. For the second step, $\sigma$ can be sampled by
standard methods for univariate sampling. In particular, provided that
the prior distribution on $\beta = 1/\sigma^2\! -\! 1$ is log-concave
(for instance true for a gamma distribution), the distribution of $\beta$ for a
known composition of the DNA mixture is also log-concave, which
means that we can use adaptive rejection sampling
\citep{gilks:wild:92} for this step.

\section{Case analysis}

We now illustrate the use of the different estimation methods for a
case analysis.  In the  Bayesian analysis, the uncertainty about
the parameters is represented by their posterior distribution. {For a
full specification of the Bayesian model we have used a uniform prior
distribution for $\theta$ and a Gamma distribution for $\beta =
1/\sigma^2\!-\!1$ with parameters $\Gamma(3.6,49)$ throughout. The chosen prior distribution for $\beta$ corresponds to a 95\% prior credibility interval of $(0.05,0.15)$ for $\sigma$, representing typical values for the variability of relative peak areas. In a specific application it would be appropriate to use a prior distribution reflecting the typical values obtained in a given forensic laboratory.  Generally it is difficult to identify an informed prior for $\theta$, as this would depend strongly on the type of trace analysed, so the uniform prior seems most appropriate.}

{Although the Bayesian method does not involve estimation of parameters but rather integrates over these, it is interesting to compare posterior means and credibility intervals to corresponding quantities when using maximum likelihood. These are displayed for illustration in Table~\ref{tab:bayesresults} for the case of one known potential contributor only.}
\begin{table}

  \caption{\label{tab:bayesresults} {Bayesian estimates (posterior means) of the mixture ratio and peak imbalance with estimated 99\% credibility intervals for the case of one known potential contributor. }}
 
 \vspace{\baselineskip}
  \centering
  
{\footnotesize
  \begin{tabular}{lcccc}
    & $\bar\sigma$ & 99\% CrI & $\bar\theta$ & 99\% CI \\ 
    \hline
    Perlin data & 0.073 & (0.052, 0.106) & 0.695 & (0.67, 0.73) \\ 
    Evett data & 0.095 & (0.063, 0.152) & 0.894 & (0.86, 0.93) \\   
    \hline
  \end{tabular}
}  
\end{table}
{The Bayesian estimates and intervals are similar to those in Table~\ref{tab:mleresults}, but the estimate and credibility interval for the peak imbalance is shifted slightly to the right by the prior information. Also the posterior correlations between $\sigma$ and $\theta$ are similar to those obtained from the information matrix, $-0.034$ for the Perlin data and $-0.166$  for the Evett data.}
\subsection{Evidence calculation}
  
\subsubsection{Evett data}
  
For the Evett data we use the known major contributor as a potential
suspect and compare the hypotheses as in \eqref{eq:LR}. Using the
fitted parameters from Table~\ref{tab:mleresults} we find
\[
\log_{10}LR = \log_{10}\frac{\den(\bR\cd H_p,\hat\sigma,
  \hat\theta)}{\den(\bR\cd H_d,\hat\sigma, \hat\theta)} =
8.53414.
\]
We have here used the MLE for the situation where the known profile is
specified to be a potential contributor. Alternatively one could use a different MLE in numerator and denominator, corresponding to the different hypotheses considered.

\begin{figure}[htb]
    \centering
    \includegraphics{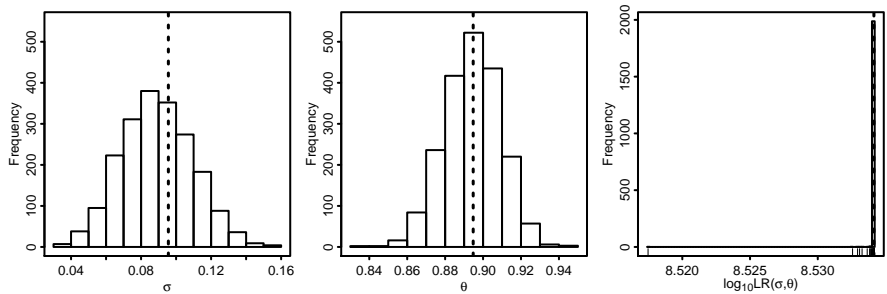}
    \caption{Bootstrap simulations of $\hat\sigma$, $\hat\theta$ and $\log_{10}LR$
   for the Evett data, based on 2000 samples. 
      The dashed lines indicate the values as estimated on the Evett data set.
    }
 \label{fig:EvettLRhists}
  \end{figure}

  As the likelihood ratio is calculated using the parameter estimates,
  we can assess the uncertainty of the estimate of $\log_{10}LR$ by
  parametric bootstrap: using the fitted parameters and the fact that
  relative sizes are Dirichlet distributed, we simulate 2000 new sets
  of relative peak sizes and estimate the parameters for each of
  these.  A 99\% bootstrap confidence interval for the
  $\log_{10}LR$ is then (8.53397, 8.53414). Note that this is very
  narrow, indicating that $\log_{10}LR$ is very accurately determined despite the uncertainty in $\sigma$ and $\theta$.  Histograms of
  the bootstrap samples of parameter estimates and $\log_{10}LR$-values are displayed in
  Figure~\ref{fig:EvettLRhists}. 
  The shape of the histograms indicate that the distribution of the estimates are well approximated by a Gaussian distribution. Note the very concentrated histogram for $\log_{10}LR$.  

  For the Bayesian analysis, the quantity of interest is the ratio of marginal
  likelihoods 
  $$LR=\frac{\den(\bR \cd H_p)}{\den(\bR \cd H_d)}$$ with both $\theta$ and $\sigma$ 
  integrated out; the numerator and denominator can be estimated from the Monte Carlo samples
  as
\[
 \den(\bR \cd H_p) \approx \frac{1}{N}\sum_{i=1}^N\den(\bR \cd H_p,\sigma_i)
\]
and similarly for $\den(\bR \cd H_d)$. This
 yields a $\log_{10}LR$ of 8.233, somewhat smaller than the value obtained by using maximum likelihood, but still representing extremely strong evidence that the suspect has contributed to the mixture. 
 
\begin{figure}[htb]
    \centering
    \includegraphics{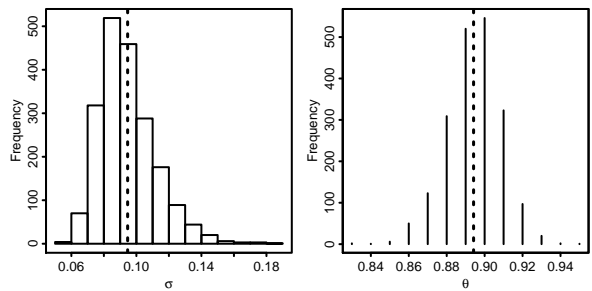}
    \caption{{Samples from the posterior distributions of  $\sigma$ and $\theta$  
   for the Evett data. 
      The dashed lines indicate the posterior mean.
    }}
 \label{fig:EvettGibbsHists}
  \end{figure}

{The marginal likelihood ratio in the Bayesian setup does not have a variation but we can compare the posterior distribution of the parameters with the bootstrap distribution of their estimates in Figure~\ref{fig:EvettLRhists}. These are shown in Figure~\ref{fig:EvettGibbsHists}. Apart from the discretization of $\theta$ in the Bayesian model, they again identify about the same region of plausible values for the parameters. }

\subsubsection{Perlin data}
\begin{figure}[htb]
    \centering
    \includegraphics{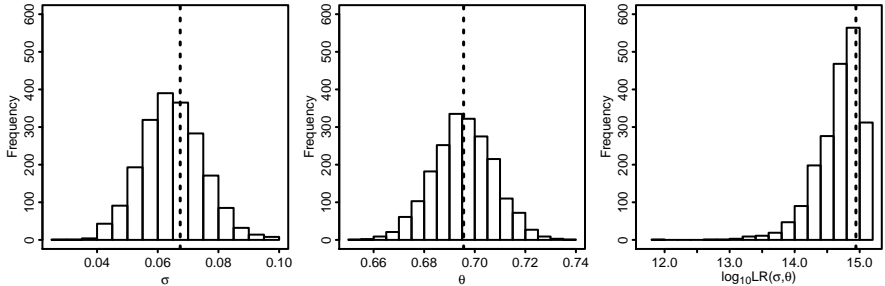}
    \caption{Bootstrap simulations of $\hat\sigma$, $\hat\theta$ and
      corresponding $\log_{10}LR(\sigma, \theta)$ for the Perlin data.
      The dashed lines indicate the values estimated from the Perlin
      data using the minor contributor as a potential contributor.}
    \label{fig:PerlinLRhists}
  \end{figure}

\begin{figure}[htb]
    \centering
    \includegraphics{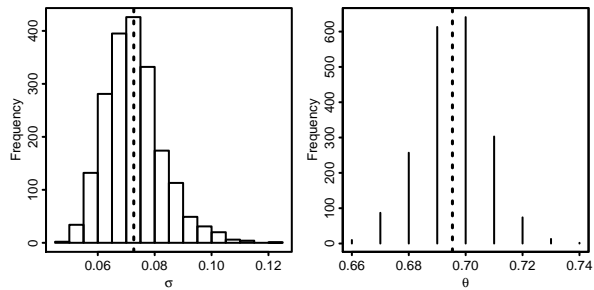}
    \caption{{Samples from the posterior distributions of $\sigma$ and $\theta$ 
   for the Perlin data. 
      The dashed lines indicate the posterior mean.}}
    \label{fig:PerlinGibbshists}
  \end{figure}

For the Perlin data, we use the known minor contributor as a potential
contributor, and consider the likelihood ratio for $H_p$ against $H_d$
resulting in $\log_{10}LR = 14.942$ using the joint maximum likelihood
estimate $(\hat\theta, \hat\sigma)$ with a 99\% bootstrap confidence
interval of (13.328, 15.075), i.e.\ a considerably wider interval than
for the Evett data. The Monte Carlo estimate for $\log_{10}LR$ of
the marginal likelihood ratio is 14.511. We have displayed
histograms for bootstrapped parameter estimates and log-likelihood
ratios in Figure~\ref{fig:PerlinLRhists} and histograms for the posterior distribution in Figure~\ref{fig:PerlinGibbshists}.

  Again, the histograms indicate that the sampling distributions of the estimates and posterior distributions  are similar and reasonably approximated by a Gaussian distribution. Here the sampling distribution of 
  $\log_{10}LR$ is more variable, indicating higher sensitivity to
  parameter uncertainty than for the Evett data. Still, all values of
  $\log_{10}LR$ in the confidence interval provide evidence that the
  potential contributor indeed did contribute to the mixture.

\subsection{Mixture deconvolution}

We produce a ranked list of probable profile pairs using the following
trick, exploiting the fact that sampling a set of DNA profiles for the
contributors is straightforward regardless of the choice of
estimation method. We sample profile pairs $(\bc_1,\bc_2)$ from a DNA
mixture model using the observed relative peak sizes and the
estimation method of preference. For each of the sampled profile pairs
we then calculate the probability of that particular pair,
$\Pro\{(\bc_1,\bc_2)\cd\bR\}$. Adding up these probabilities for all
the sampled profile pairs we get that they account for some total $p$
of the probability mass, implying that no undiscovered pair of
profiles can have probability larger than $1-p$. Thus, if $k$ of the
sampled profile-pairs all have probability larger than $1-p$ they must
constitute the $k$ most probable profiles. Note that the number $k$ of
most probable profiles is not fixed in advance, and that increasing
the number of samples will result in a longer list of 
profiles. We illustrate this method using the Perlin data.

If the model is fitted using maximum likelihood estimates for $\sigma$
and $\theta$, we can sample possible profiles for the two contributors
using the Bayesian network and thereby obtain pairs of profiles of
high probability.  The sampling revealed 13 different profiles with a
total probability of $p=0.9992$.  There were eight of these profiles
with a probability larger than $1-p=0.0008$, implying
that the $k=8$ most probable DNA profile pairs
had been determined. For seven of the markers the genotypes were correctly
identified; for the marker TH01 there is slight uncertainty whether
the minor contributor supplied the allele 7 or 9; similarly, for
markers D19 and VWA there is slight uncertainty about the allocation
of alleles to the contributors.  Table~\ref{tab:PerlinsepMLE} displays
the eight possible choices for the remaining three markers.

\begin{table}
 \caption{The eight most probable contributor genotypes of the three uncertain markers for the Perlin data using the MLE for $\sigma$ and $\theta$. Correctly predicted genotypes are marked in bold.}
\label{tab:PerlinsepMLE}

\vspace{\baselineskip}
    \centering \small
\begin{tabular}{cccccccc}
&\multicolumn{3}{c}{Major contributor} & \multicolumn{3}{c}{Minor contributor} & \\
rank&D19 & TH01 & VWA & D19 & TH01 & VWA & Prob. \\ 
  \hline
1&\bf{12.2 15} & \bf{7 9} & \bf{17 17} & \bf{14 14} & \bf{6 7} & \bf{18 18} & 0.647 \\ 
  2&\bf{12.2 15} & \bf{7 9} & 17 18 & \bf{14 14} & \bf{6 7} & 17 17 & 0.261 \\ 
  3&12.2 14 & \bf{7 9} & \bf{17 17} & 15 15 & \bf{6 7} & \bf{18 18} & 0.054 \\ 
  4&12.2 14 & \bf{7 9} & 17 18 & 15 15 & \bf{6 7} & 17 17 & 0.022 \\ 
 5& 14 15 & \bf{7 9} & \bf{17 17} & 12.2 12.2 & \bf{6 7} & \bf{18 18} & 0.006 \\ 
 6& \bf{12.2 15} & \bf{7 9} & \bf{17 17} & \bf{14 14} & 6 9 & \bf{18 18} & 0.004 \\ 
 7& 14 15 & \bf{7 9} & 17 18 & 12.2 12.2 & \bf{6 7} & 17 17 & 0.002 \\ 
 8& \bf{12.2 15} & \bf{7 9} & 17 18 & \bf{14 14} & 6 9 & 17 17 & 0.001 \\ 
   \hline
   \multicolumn{7}{r}{Total probability} & 0.997 \\
\hline
\end{tabular}
\end{table}

We note that for the Perlin data the most probable profile pair is
 the true one and the second most probable pair has a
misclassification on only one marker, VWA. As for the analysis of
evidential value, it is possible to assess the uncertainty of these
rankings and the sensitivity to the choice of parameters, e.g. by
bootstrap. We shall omit such further analysis here.

For the Bayesian analysis 
we use the Gibbs sampler to locate high probability pairs of profiles.
In this case we obtain 27 
different profiles for each contributor.
Subsequently we again use the Gibbs sampler to obtain the posterior
probability $\Pro\{(\bc_1,\bc_2)\cd \bR\}$ for each pair as

\[
\Pro\{(\bc_1,\bc_2)\cd \bR\} \approx \frac{1}{N}\sum_{i=1}^N\Pro\{(\bc_1,\bc_2) \cd \bR, \sigma_i\}.
\] 
This yields a total probability $p=0.998$ for the 27 profiles and
identifies nine profiles with a probability larger than the resulting
threshold $1-p=0.002$. 
Also in the Bayesian analysis all genotypes were correctly identified
for seven of the markers. The genotypes of the profile pairs for the
remaining three markers are displayed in
Table~\ref{tab:PerlinsepMCMC}.

\begin{table}
  \caption{The nine 
    most probable contributor genotypes of the three uncertain markers for the Perlin data in a Bayesian analysis.}
  \label{tab:PerlinsepMCMC}
    \centering \small   
   \vspace{\baselineskip}
   \begin{tabular}{cccccccc}
&\multicolumn{3}{c}{Major contributor} & \multicolumn{3}{c}{Minor contributor} & \\
rank&D19 & TH01 & VWA & D19 & TH01 & VWA & Prob. \\ 
  \hline
1 & \bf{12.2 15} & \bf{7 9} & \bf{17 17} & \bf{14 14} & \bf{6 7} & \bf{18 18} & 0.548 \\ 
  2 & \bf{12.2 15} & \bf{7 9} & 17 18 & \bf{14 14} & \bf{6 7} & 17 17 & 0.300 \\ 
  3 & 12.2 14 & \bf{7 9} & 17 18 & 15 15 & \bf{6 7} & 17 17 & 0.061 \\ 
  4 & 12.2 14 & \bf{7 9} & \bf{17 17} & 15 15 & \bf{6 7} & \bf{18 18} & 0.055 \\ 
  5 & \bf{12.2 15} & \bf{7 9} & \bf{17 17} & \bf{14 14} & 6 9 & \bf{18 18} & 0.007 \\ 
  6 & \bf{12.2 15} & \bf{7 9} & 17 18 & \bf{14 14} & 6 9 & 17 17 & 0.005 \\ 
  7 & \bf{12.2 15} & \bf{7 9} & \bf{17 17} & \bf{14 14} & \bf{6 7} & 17 18 & 0.005 \\ 
  8 & 14 15 & \bf{7 9} & \bf{17 17} & 12.2 12.2 & \bf{6 7} & \bf{18 18} & 0.005 \\ 
  9 & 14 15 & \bf{7 9} & 17 18 & 12.2 12.2 & \bf{6 7} & 17 17 & 0.003 \\ 
   \hline
   \multicolumn{7}{r}{Total probability} & 0.988 \\
\hline

\end{tabular}
\end{table}

The Bayesian analysis reveals the same two profile pairs as the most
probable but slightly reverses the ranking of profile pairs with
smaller probabilities. Note, though, that the probabilities in
Table~\ref{tab:PerlinsepMCMC} are Monte Carlo estimates with a
standard error up to 3\% and thus the mutual ranking of profiles with
very similar probabilities is uncertain. In other words, the four most
highly ranked configurations are definitely the top four on the list,
whereas the mutual ranking of number three and four could be reversed,
although this is unlikely. Similarly, the correct mutual ranking of
the last five could easily be reversed in comparison
with the ranking in the table.
 
\section{Discussion}
In the present article we have demonstrated how both the mixture
proportions and the unknown variance factor in the model used by
\cite{article:Gammamodel} can be estimated and its uncertainty
incorporated into further analysis of the DNA trace. The analysis
shows that there is sufficient information in a single trace to do so
using only the peak sizes for the data at hand, if the variance factor
$\sigma$ is taken to be marker independent.

We have illustrated how it is possible to assess the
performance of a particular method for a given case. It would be well
worthwhile to carry out a further study on the general performance,
for example of the stability of the method for mixture
deconvolution. However, we would like to emphasise that by performing
a bootstrap analysis we get an indication of the information that data
from mixtures of similar composition would provide about the questions
in mind and the bootstrap analyses do indicate a considerable
stability of the findings.

There might be good reasons to believe that the peak imbalance
$\sigma$ differs across markers. As there is limited information in
the data from a single case we have for practical reasons chosen to
ignore this variation and use a single $\sigma$ for each case. An
alternative way of accommodating marker dependence on the generic
variability would be to assume that the parameter $\beta =
1/\sigma^2\!-\!1$ in the gamma model (\ref{eq:gamma}) depends on $m$
as $\beta_m=\lambda\delta_m$ where $\lambda$ depends only on the case
at hand and $\delta_m$ is marker dependent but independent of the case
considered. One could then use laboratory data to estimate $\delta_m$
and only adapt $\lambda$ to the case. This methodology
would only demand minor technical variations for the Bayesian and
maximum likelihood methods developed in this paper.  Generally, prior
information on the total amount of DNA would also be available which
could be used to improve the Bayesian analysis by using an informed
prior distribution for $\lambda$.

We have only considered the simplest model with a fixed number of
contributors and no artefacts in the form of stutter, dropout, etc.
Maximum likelihood estimates for the parameters can still be found
extending to the case of multiple contributors, but as there would be
more mixture proportions to estimate, larger confidence intervals for
the parameter estimates are to be expected. Extensions \citep{cowell:etal:11,cowell:etal:13} involve even more
parameters which need to be estimated in a similar way and it
certainly adds to the general complexity of the problem, as does
issues of gene frequency uncertainties etc.\ \citep{green:mortera:09}.
We expect to address these issues in the future.

\section*{Acknowledgements} We are grateful to Kjell Konis for advice
and flexibility in adapting \texttt{RHugin} to serve the purpose of
this analysis, and to Robert Cowell and Julia Mortera
for helpful
comments on a previous version of the manuscript.


\clearpage
\begin{thebibliography}{23}
\expandafter\ifx\csname natexlab\endcsname\relax\def\natexlab#1{#1}\fi
\expandafter\ifx\csname url\endcsname\relax
  \def\url#1{\texttt{#1}}\fi
\expandafter\ifx\csname urlprefix\endcsname\relax\def\urlprefix{URL }\fi

\bibitem[{Bill et~al.(2005)Bill, Gill, Curran, Clayton, Pinchin, Healy, and
  Buckleton}]{article:PENDULUM}
Bill, M., Gill, P., Curran, J., Clayton, T., Pinchin, R., Healy, M., Buckleton,
  J., 2005. {PENDULUM} -- a guideline-based approach to the interpretation of
  {STR} mixtures. Forensic Science International 148, 181--189.

\bibitem[{Butler(2005)}]{book:Butler}
Butler, J.~M., February 2005. Forensic {DNA} Typing: Biology, Technology, and
  Genetics of {STR} Markers, 2nd Edition. Elsevier ACADEMIC PRESS.

\bibitem[{Butler et~al.(2003)Butler, Schoske, Vallone, Redman, and
  Kline}]{butler:etal:03}
Butler, J.~M., Schoske, R., Vallone, P.~M., Redman, J.~W., Kline, M.~C., 2003.
  Allele frequencies for 15 autosomal loci on {U.S.} {C}aucasian, {A}frican
  {A}merican, and {H}ispanic populations. Journal of Forensic Science 48~(4).

\bibitem[{Clayton et~al.(1998)Clayton, Whitaker, Sparkes, and
  Gill}]{Clayton199855}
Clayton, T., Whitaker, J., Sparkes, R., Gill, P., 1998. Analysis and
  interpretation of mixed forensic stains using {DNA} {STR} profiling. Forensic
  Science International 91~(1), 55--70.

\bibitem[{Cowell(2009)}]{Cowell2009193}
Cowell, R.~G., 2009. Validation of an {STR} peak area model. Forensic Science
  International: Genetics 3~(3), 193--199.

\bibitem[{Cowell et~al.(2013)Cowell, Graversen, Lauritzen, and
  Mortera}]{cowell:etal:13}
Cowell, R.~G., Graversen, T., Lauritzen, S., Mortera, J., 2013. Analysis of
  {DNA} mixtures with artefacts, arXiv:1302:4404.

\bibitem[{Cowell et~al.(2007{\natexlab{a}})Cowell, Lauritzen, and
  Mortera}]{article:Gammamodel}
Cowell, R.~G., Lauritzen, S.~L., Mortera, J., January 2007{\natexlab{a}}. A
  gamma model for {DNA} mixture analyses. Bayesian Analysis 2~(2), 333--348.

\bibitem[{Cowell et~al.(2007{\natexlab{b}})Cowell, Lauritzen, and
  Mortera}]{rgc/sll/jm:fsi}
Cowell, R.~G., Lauritzen, S.~L., Mortera, J., 2007{\natexlab{b}}.
  Identification and separation of {DNA} mixtures using peak area information.
  Forensic Science International 166, 28--34.

\bibitem[{Cowell et~al.(2011)Cowell, Lauritzen, and Mortera}]{cowell:etal:11}
Cowell, R.~G., Lauritzen, S.~L., Mortera, J., 2011. Probabilistic expert
  systems for handling artifacts in complex {DNA} mixtures. Forensic Science
  International: Genetics 5, 202--209.

\bibitem[{Curran(2008)}]{CurranMCMC}
Curran, J.~M., 2008. A {MCMC} method for resolving two person mixtures. Science
  and Justice 48, 168--177.

\bibitem[{Evett et~al.(1998)Evett, Gill, and Lambert}]{article:Evettdata}
Evett, I., Gill, P., Lambert, J., 1998. Taking account of peak areas when
  interpreting mixed {DNA} profiles. Journal of Forensic Science 43~(1),
  62--69.

\bibitem[{Gilks et~al.(1996)Gilks, Richardson, and
  Spiegelhalter}]{gilks:richardson:spiegelhalter:96a}
Gilks, W.~R., Richardson, S., Spiegelhalter, D.~J., 1996. {M}arkov {C}hain
  {M}onte {C}arlo Methods in Practice. Chapman and Hall, New York.

\bibitem[{Gilks and Wild(1992)}]{gilks:wild:92}
Gilks, W.~R., Wild, P., 1992. Adaptive rejection sampling for {G}ibbs sampling.
  Applied Statistics 41~(2), 337--348.

\bibitem[{Gill et~al.(2006)Gill, Brenner, Buckleton, Carracedo, Krawczak, Mayr,
  Morling, Prinz, Schneider, and Weir}]{Gill200690}
Gill, P., Brenner, C., Buckleton, J., Carracedo, A., Krawczak, M., Mayr, W.,
  Morling, N., Prinz, M., Schneider, P., Weir, B., 2006. {DNA} commission of
  the international society of forensic genetics: Recommendations on the
  interpretation of mixtures. Forensic Science International 160~(2), 90 --
  101.

\bibitem[{Gill et~al.(2008)Gill, Curran, Neumann, Kirkham, Clayton, Whitaker,
  and Lambert}]{Gill200891}
Gill, P., Curran, J., Neumann, C., Kirkham, A., Clayton, T., Whitaker, J.,
  Lambert, J., 2008. Interpretation of complex {DNA} profiles using empirical
  models and a method to measure their robustness. Forensic Science
  International: Genetics 2~(2), 91--103.

\bibitem[{Green and Mortera(2009)}]{green:mortera:09}
Green, P.~J., Mortera, J., 2009. Sensitivity of inferences in forensic genetics
  to assumptions about founder genes. Annals of Applied Statistics 3~(2),
  731--763.

\bibitem[{HUGIN API(2009)}]{manual:HUGINapi}
HUGIN API, 2009. {HUGIN API} Reference Manual. Hugin Expert A/S.

\bibitem[{Konis and {Hugin Expert A/S}(2010)}]{manual:RHugin}
Konis, K., {Hugin Expert A/S}, 2010. RHugin. {R} package version 7.3-1.

\bibitem[{Perlin and Szabady(2001)}]{article:Perlindata}
Perlin, M., Szabady, B., 2001. Linear mixture analysis: a mathematical approach
  to resolving mixed {DNA} samples. Journal of Forensic Science 46, 1372--1378.

\bibitem[{Perlin et~al.(2011)Perlin, Legler, Spencer, Smith, Allan, Belrose,
  and Duceman}]{perlin:2011}
Perlin, M.~W., Legler, M.~M., Spencer, C.~E., Smith, J.~L., Allan, W.~P.,
  Belrose, J.~L., Duceman, B.~W., 2011. Validating
  {TrueAllele\textsuperscript{\textregistered}} {DNA} mixture interpretation.
  Journal of Forensic Sciences 56~(6), 1430--1447.

\bibitem[{Puch-Solis et~al.(2012)Puch-Solis, Rodgers, Mazumder, Pope, Evett,
  Curran, and Balding}]{puch2012}
Puch-Solis, R., Rodgers, L., Mazumder, A., Pope, S., Evett, I., Curran, J.,
  Balding, D., 2012. Evaluating forensic {DNA} profiles using peak heights,
  allowing for multiple donors, allelic dropout and stutters. Tech. rep., LGC
  Research Report LGC/P/2012/138.

\bibitem[{{{R} Development Core Team}(2011)}]{manual:R}
{{R} Development Core Team}, 2011. R: A Language and Environment for
  Statistical Computing. {R} Foundation for Statistical Computing, Vienna,
  Austria, {ISBN} 3-900051-07-0.

\bibitem[{Wang et~al.(2006)Wang, Xue, and Birdwell}]{wang2006least}
Wang, T., Xue, N., Birdwell, J.~D., 2006. Least-square deconvolution: A
  framework for interpreting short tandem repeat mixtures. Journal of Forensic
  Sciences 51~(6), 1284--1297.

\end{thebibliography}
\end{document}